\documentstyle[12pt,epsfig,axodraw,a4]{article} 
\def\bild#1#2{    
        \vspace*{-5mm}
        \begin{center}
        \begin{math}
        \epsfxsize#2cm
        \epsffile{#1}
        \end{math}
        \end{center}  }
\newcommand{\vs}{\vspace{-0.25cm}}
\begin{document} 
\begin{center}
\large{\bf Chiral corrections to the isovector double scattering term for 
the pion-deuteron scattering length}

\medskip

N. Kaiser\\

\smallskip

{\small Physik Department T39, Technische Universit\"{a}t M\"{u}nchen,
    D-85747 Garching, Germany}

\end{center}

\bigskip

\begin{abstract}
The empirical value of the real part of the pion-deuteron scattering length can
be well understood in terms of the dominant isovector $\pi N$-double scattering
contribution. We calculate in chiral perturbation theory all one-pion loop 
corrections to this double scattering term which in the case of $\pi 
N$-scattering close the gap between the current-algebra prediction and the 
empirical value of the isovector threshold T-matrix $T_{\pi N}^-$. In addition
to closing this gap there is in the $\pi d$-system a loop-induced off-shell 
correction for the exchanged virtual pion. Its coordinate space representation
reveals that it is equivalent to $2\pi$-exchange in the deuteron. We evaluate 
the chirally corrected double scattering term and the off-shell contribution 
with various realistic deuteron wave functions. We find that the off-shell 
correction contributes at most -8\%  and that the isovector double scattering 
term explains at least 90\% of the empirical value of the real part of the 
$\pi d$-scattering length.

\end{abstract}


To be published in {\it The Physical Review C (2002), Brief Reports}

\bigskip

\vspace{3cm}

The tool to systematically investigate the consequences of spontaneous and 
explicit chiral symmetry breaking in QCD is chiral perturbation theory. 
Recently, methodology has been developed \cite{weinberg} which allows one to 
relate scattering processes involving a single nucleon to nuclear scattering 
processes. For instance, one can relate pion-nucleon scattering to 
pion-deuteron scattering. The non-perturbative effects responsible for deuteron
binding are accounted for by (phenomenological) deuteron wavefunctions. 
Although this introduces some model dependence one can evaluate matrix elements
of chiral operators in a variety of wavefunctions in order to estimate the 
theoretical error induced by the off-shell behavior of different NN-potentials.
 
In a recent precision experiment the PSI-group has extracted the isoscalar and 
isovector pion-nucleon scattering lengths $a^{\pm}_{\pi N}$ from the shift and 
the width of the 1s-level in pionic hydrogen, with the result \cite{schroeder}:
 \begin{equation} T^+_{\pi N} = 4\pi \Big( 1 +{m_\pi \over M_p}\Big)\, a^+_{\pi
N} = (-0.045 \pm 0.088)\, {\rm fm}\,,  \end{equation}
\begin{equation} T^-_{\pi N} = 4\pi \Big( 1 +{m_\pi \over M_p}\Big)\, a^-_{\pi
N} = (1.847 \mp 0.086)\, {\rm fm}\,. \end{equation}
One concludes from these results that the isoscalar $\pi N$-scattering length 
$a^+_{\pi N}$ is compatible with zero and that the (present) errors of 
$a^\pm_{\pi N}$ are anti-correlated. The aim of future PSI-experiments is to 
measure the level width $\Gamma_{1s} \sim a^-_{\pi N}$ with the same precision 
as the level shift $\epsilon_{1s} \sim  a^+_{\pi N}+a^-_{\pi N}$. We have 
converted in eqs.(1,2) scattering lengths into equivalent threshold T-matrices 
by multiplying with the appropriate kinematical factor. $M_p = 938.27$\,MeV and
$m_\pi=139.57\, $MeV denote the proton and charged pion mass. 

In an analogous PSI-experiment the complex-valued pion-deuteron scattering 
length $a_{\pi d}$ has been extracted from the shift and the width of the 
1s-level in pionic deuterium \cite{chat,hauser}. Again converted into a 
threshold T-matrix the (averaged) experimental result of 
refs.\cite{chat,hauser} reads for the real part, 
\begin{equation} {\rm Re}\, T_{\pi d} = 4\pi \Big( 1 +{m_\pi \over M_d}
\Big)\, {\rm Re}\, a_{\pi d} = (-0.496 \pm 0.016)\, {\rm fm}\,, \end{equation} 
where $M_d=1875.6$\,MeV denotes the deuteron mass. 

Because of the large size (and weak binding) of the deuteron it is possible to
establish an accurate relationship \cite{erics} between the real part 
Re\,$ T_{\pi d}$ and $\pi N$ threshold T-matrices $T^\pm_{\pi N}$. The single 
scattering contribution $ T^{(s)}_{\pi d}= 2T^+_{\pi N}$ is small and within
present error bars compatible with zero. The empirical value of the real part 
Re\,$T_{\pi d}$ can be well understood in terms of the dominant isovector
double scattering term \cite{erics} which reads,  
\begin{equation}  T^{(d)}_{\pi d} = -\Big(T^-_{\pi N} \Big)^2 
\int_0^\infty dr\, {\rho(r) \over \pi r} \,, \qquad \rho(r)=u^2(r)+w^2(r) \,. 
\end{equation}
Here, $u(r)$ and $w(r)$ denote the conventional s- and d-state deuteron
wavefunctions, normalized to $\int_0^\infty dr \, \rho(r)=1$. The relation
eq.(4) can be easily derived from the diagrams shown in Fig.1. The 
four-momentum of the in- and out-going pion (marked by arrows) is $(m_\pi,
\vec 0\,)$ and that of the exchanged virtual pion is $(m_\pi,\vec q\,)$. After
Fourier-transforming the product of the corresponding  pion-propagator $1/\vec 
q\,^2$ and the deuteron momentum-distributions into coordinate space one gets 
$1/(4\pi r)$ times the deuteron density $\rho(r)$. The remaining factor $-4$ 
originates from the isospin factors of the diagrams. The advantage of working 
with threshold T-matrices (which are directly related to Feynman diagrams) is 
that there appear no additional kinematical factors in eq.(4). Other three-body
interaction terms induced by $\pi\pi$-interaction (the in- and out-going pion
couple to a virtual pion in flight) which are formally of the same chiral order
have been evaluated in refs.\cite{weinberg,beane} and they were found to be
negligibly small. For extensive earlier works on the pion-deuteron scattering
length and related issues, see
refs.\cite{kolybasov,myhrer,afnan,faeldt,landau,baru}. 

The shortcoming of the present chiral 
perturbation theory calculations of Re\,$T_{\pi d}$ is that the isovector 
double scattering contribution eq.(4) is (and for reasons of consistency has 
to be) evaluated with the leading order expression $T^-_{\pi N}({\rm tree})=
m_\pi/2 f_\pi^2= 1.61\,$fm for the isovector $\pi N$ threshold T-matrix. As
usual $f_\pi = 92.4\,$MeV denotes the weak pion decay constant. Since this
Tomozawa-Weinberg prediction $T^-_{\pi N}({\rm tree})=m_\pi/2f_\pi^2$ is about
13\% smaller than the empirical (central)  value of $T^-_{\pi N}$ (see eq.(2))
the (leading order) chiral isovector double scattering contribution to Re\,$
T_{\pi d}$ comes out about 25\% too small, as can be seen from the numbers
given in Table\,1 of ref.\cite{beane}.  In fact, Weinberg argued in
ref.\cite{weinberg} that because of its dominance the isovector double
scattering contribution eq.(4) should be calculated including vertex
corrections and he simply took the measured  $\pi N$-scattering length
$a^-_{\pi N}$. The purpose of the present paper is to study in the systematic
framework of chiral perturbation theory the corrections to this simple
substitution rule.    


\bigskip

\bild{deutfig1.epsi}{14}
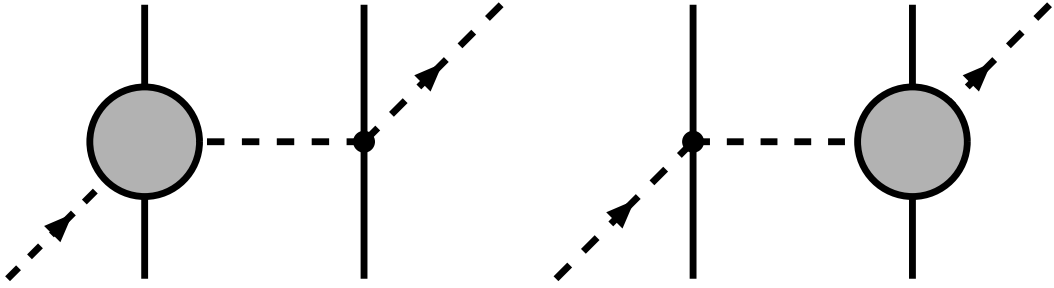
{\it Fig.\,1: Isovector double scattering contributions to pion-deuteron 
scattering at threshold. The grey disk symbolizes all one-loop graphs of 
isovector $\pi N$-scattering. Diagrams for which the role of the in- and 
out-going pion is interchanged are not shown.}

\bigskip


It has been shown in ref.\cite{pinloop} that the about 13\% gap between the 
current algebra prediction $T^-_{\pi N}({\rm tree})=m_\pi/2f_\pi^2$ and the 
empirical value of 
$T^-_{\pi N}$ is closed by chiral one-pion loop corrections at order ${\cal
O}(m_\pi^3)$. All other possible corrections from resonances etc. are much too
small to close this gap. Keeping only the important pion-loop correction the
chiral expansion of the on-shell isovector $\pi N$ threshold T-matrix reads
\cite{pinloop},   
\begin{equation} T^-_{\pi N}= {m_\pi \over 2f_\pi^2} + {m_\pi^3 
\over 16 \pi^2 f_\pi^4} \Big( 1-2 \ln{m_\pi \over \lambda} \Big) \,,
\end{equation}
with $\lambda$ a scale parameter introduced in dimensional regularization and 
minimal subtraction. With $\lambda = 1.04\,$GeV the central empirical value of
$T^-_{\pi N}$ is reproduced. The scale parameter $\lambda$ can also be 
interpreted as a momentum space cut-off $\Lambda$. By comparing with the 
expression of relevant (logarithmically divergent) loop integral obtained in 
cut-off regularization one finds the relation $\Lambda =  \sqrt{e} \,\lambda/2=
857\,$MeV. This value of the cut-off $\Lambda$ is comparable to the usual
estimate of the chiral symmetry breaking scale $\Lambda_\chi = 2\sqrt{2} \pi
f_\pi \simeq 820\,$MeV.   

In order to improve on the isovector double scattering contribution $T^{(d)
}_{\pi d}$ one should therefore include pion-loop corrections at the $\pi N\to 
\pi N$ transition vertices. The grey disk in the diagrams in Fig.\,1 symbolizes
all one-loop graphs of (isovector) $\pi N$-scattering. Now, since the exchanged
pion with four-momentum $(m_\pi,\vec q\,)$ is off its mass-shell the loop 
diagrams included in the grey disk will lead for such a 
half off-shell kinematics to more than the order ${\cal O}(m_\pi^3)$ loop
correction to $T^-_{\pi N}$ given in eq.(5).

\bigskip

\bild{deutfig2.epsi}{9}
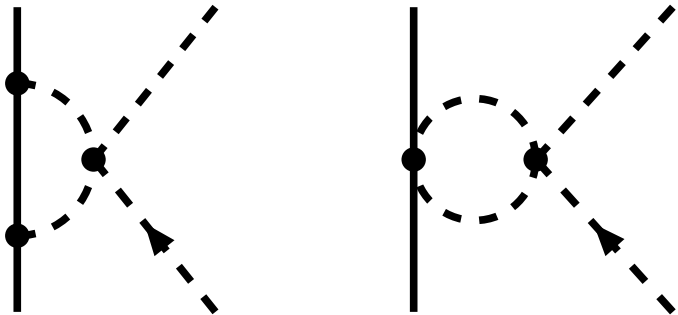
{\it Fig.\,1: Those one-loop graphs of isovector $\pi N$-scattering which
generate the off-shell correction $\delta T^-_{\pi N}$(off) for the exchanged 
virtual pion in the $\pi d$-system.}

\bigskip


In addition there is an off-shell contribution $\delta T^-_{\pi N}$(off) which 
turns out to be generated entirely by the two graphs shown in Fig.\,2 
involving the chiral $4\pi$-vertex. This off-shell contribution reads
explicitly, 
\begin{eqnarray}  \delta T^-_{\pi N}({\rm off}) &=&{m_\pi\over96\pi^2 f_\pi^4} 
\bigg\{ q^2 (1+5g_A^2)\ln{m_\pi \over \lambda}+{q^2 \over 6} (1+17g_A^2)
\nonumber \\ && + \Big[4m_\pi^2(1+2g_A^2) +q^2(1+5g_A^2) \Big]  \bigg[
{s \over q} \ln {s+ q \over 2m_\pi} -1 \bigg] \bigg\} \nonumber \\ &=&  {m_\pi 
\over 2f_\pi^2} \,G^V_E(-q^2 )_{loop}\,, \end{eqnarray}
with the abbreviation $s=\sqrt{4m_\pi^2+q^2}$. As indicated it is proportional 
to the one-pion loop contribution to the isovector electric form factor of the 
nucleon $G^V_E(-q^2)$ (normalized to unity at $q^2=0$). We also note that the 
off-shell correction $\delta T^-_{\pi N}$(off) in eq.(6) is independent of the
choice of the interpolating pion-field even though the (off-shell)
$4\pi$-vertex is not.

With inclusion of chiral pion-loop corrections to the threshold T-matrix 
$T^-_{\pi N}$ the isovector double scattering term takes the form,
\begin{equation} T^{(d)}_{\pi d}={m_\pi^2 \over 4f_\pi^4 }\bigg[ -1+
{m_\pi^2 \over 2\pi^2 f_\pi^2} \ln{m_\pi \over 2\Lambda} \bigg] \int_{1/
\Lambda}^\infty dr\, {\rho(r) \over \pi r} \,,  \end{equation}
which is practically equivalent to inserting the empirical value of $T^-_{\pi
N}$ into eq.(4).  The effect of the off-shell correction eq.(6) to Re\,$T_{\pi
d}$ is obtained by multiplying $\delta T^-_{\pi N}$(off) with the 
pion-propagator $1/\vec q\,^2$ and Fourier-transforming this product into
coordinate  space. The terms in the first line of eq.(6) give this way a
delta-function $\delta^3(\vec r\,)$ which exclusively probes the deuteron 
wavefunction at the origin. The Fourier-transform of the other non-polynomial 
terms in eq.(6) is most conveniently obtained with the help of a dispersion
relation and one  finds
\begin{equation} \delta T^{(o)}_{\pi d} = {m_\pi^2 \over 96\pi^3 f_\pi^6}
\int_{1/\Lambda}^\infty dr \, {\rho(r)\over r}  \int_{2m_\pi}^\infty 
d\mu \, e^{-\mu r} \sqrt{\mu^2-4m_\pi^2} \,\bigg[ 1+ 5g_A^2 -{4m_\pi^2 \over
\mu^2} (1+2g_A^2) \bigg] \,. \end{equation} 
One observes that the weight function multiplying the deuteron density
$\rho(r)$ goes like $r^{-3}$ for $r\to 0$. On the other hand its asymptotic 
behavior for $r\to \infty$ is $e^{-2m_\pi r} r^{-5/2}$, the typical form of a
$2\pi$-exchange potential. In contrast to the double scattering term eq.(7) the
off-shell correction eq.(8) is mainly sensitive to the short distance behavior
of the deuteron wavefunction. The $r^{-3}$-singularity in eq.(8) originates 
from the large-$q$ behavior of the (one-loop) form factor in eq.(6), which
clearly is not realistic for large momentum transfers. We regularize the
$r^{-3}$ short distance singularity (in a minimal way) by starting the radial
integration at a nonzero $r_{min} = 1/\Lambda  = 0.23\,$fm, where $\Lambda
=857\,$MeV is the (equivalent) momentum cut-off entering the chiral logarithm
in eq.(5).  
\vspace{-0.4cm}

\begin{table}[hbt]
\begin{center}
\begin{tabular}{|c|ccccccc|}
    \hline
    Potential & Paris\cite{paris} & Bonn-CD\cite{cdbonn} & Idaho\cite{idaho}& 
NL5\cite{evgeni} & NL6\cite{evgeni} & NNL5\cite{evgeni} & NNL6\cite{evgeni} 
\\\hline  $T^{(d)}_{\pi d}$ [fm]
& $-$0.479 &$-$0.495 &$-$0.466 & $-$0.478 & $-$0.459 &$-$0.500 &$-$0.499 \\ 
 $\delta T^{(o)}_{\pi d} $ [fm]
& 0.015 &0.025 &0.017 &0.023& 0.008& 0.036 &0.036 \\ 
  \hline
  \end{tabular}
\end{center}
\end{table}
\vspace{-0.5cm}
{\it Tab.1: Numerical values of the chirally corrected double scattering
contribution $T^{(d)}_{\pi d}$ and the off-shell correction $\delta
T^{(o)}_{\pi d} $ for various deuteron wavefunctions.}

\bigskip

In Table\, 1, we present numerical values of the chirally corrected double 
scattering contribution $T^{(d)}_{\pi d}$ and the off-shell correction 
$\delta T^{(o)}_{\pi d} $ for various realistic deuteron wavefunctions taken
from refs.\cite{paris,cdbonn,idaho,evgeni}. One observes that the chirally 
corrected isovector double scattering contribution $T^{(d)}_{\pi d}$ is
rather stable and close to the empirical value of Re\,$T_{\pi d}$ in eq.(3). 
When starting the radial integration at $r=0$ instead of $r_{min}=0.23\,$fm 
these numbers are affected at most in the last digit. The off-shell correction
$\delta T^{(o)}_{\pi d}$ which is mainly sensitive to the short distance
behavior of $\rho(r)$ varies more with the deuteron wavefunction. It is however
always a small attractive effect whose relative magnitude does not exceed 8\%. 

We conclude that the (chirally corrected) isovector double scattering term 
explains at least 90\% of the empirical value of the real part of the $\pi d$ 
threshold T-matrix. This is in agreement with the findings of ref.\cite{tony} 
were a host of other small corrections to Re\,$T_{\pi d}$ has been 
investigated. For example, in ref.\cite{tony} an off-shell correction of the 
form eq.(6) proportional to a phenomenological dipole form factor (minus unity)
has been considered. The off-shell correction eq.(6) obtained in the present
work has in fact a well-defined conceptual and physical origin. It is the
unique byproduct of the pion-loop graphs which close the gap between the
current-algebra prediction and the empirical value of the isovector $\pi N$
threshold T-matrix $T_{\pi N}$. Its effect on Re\,$T_{\pi d}$ is comparable to
the inherent theoretical uncertainty of the isovector double scattering
contribution $T^{(d)}_{\pi d}$.  
\section*{Acknowledgements}
I thank E. Epelbaum and R. Machleidt for providing me tabulated deuteron
wavefunctions.

\end{document}